\documentclass[aps,prl,amsmath,amssymb,nofootinbib,superscriptaddress,twocolumn,10pt]{revtex4-2}
\usepackage{bm,psfrag,graphicx,setspace}
\usepackage[T1]{fontenc}
\usepackage[left]{lineno}
\usepackage{xcolor}

\usepackage{bbold}
\usepackage{mathtools}
\usepackage{braket}
\DeclareRobustCommand{\igmas}{\text{\reflectbox{$\sigma$}}}

\selectfont
\begin{document}
\bibliographystyle{apsrev4-2}
\author{Kevin N. Moser}
\author{Marc R. Bourgeois}
\author{David J. Masiello}
\email{masiello@uw.edu}
\affiliation{Department of Chemistry, University of Washington, Seattle, Washington 98195, USA}
%%%%%%%%%%%%%%%%%%%%%%%%%%%%%%%%%%%%%%%%
\begin{abstract}
Signatures of a material object's broken inversion symmetry within its intrinsic excitations and associated optical fields have largely evaded characterization from a perspective that evenly accounts for the inseparability between matter and field dynamics. Here, we formulate a complex-valued pseudoscalar chirality metric that is derived from the coupled electromagnetic and material governing equations, and demonstrate its resolution of the excitational chirality exhibited by the eigenmodes and eigenfields of a family of structurally chiral objects, including those with chirally connected enantiomeric states.
\end{abstract}

\title{Unified light-matter metric of molecular and nanophotonic chirality}

\maketitle

The nonsuperimposability of a chiral object's static geometry with that of its mirror image has profound influence upon the emergent structure of matter so diverse as the homochiral building blocks of life \cite{young2019mirror,xue2018enzymatic,harrison2023synthesis,blackmond2010origin} and the spiraling patterns of distant galaxies \cite{PhysRevLett.124.101302,capozziello2006spiral}. Yet, no matter is truly quiescent. Even in the absence of all external forces, and at zero temperature, it undulates incessantly at its intrinsic characteristic frequencies. Any participating charges command the appearance of companion fields, as dictated by the correspondence principle, which encode information on the sourcing object's intrinsic excitations and underlying static structure in their evanescent and propagating components.

Beyond the binary presence or absence of parity symmetry, quantification of the degree of structural \cite{cotton} and dynamic \cite{barron1986true} chirality in molecular and nanophotonic systems has driven the invention of a menagerie of metrics based on consideration of mathematical constructions \cite{zabrodsky1995continuous, cha2024graph, feng2025chiral,osipov1995new} or physical principles \cite{PhysRevX.6.031013,vavilin2022multidimensional,bradshaw2015signatures}, the latter being extended to examine intrinsic material excitations through use of generalized helicity measures \cite{ PhysRevLett.133.268001, 8f2s-rjgy,bezard2026absolute,mackinnon2025quantized}. Meanwhile, the local handedness of the electromagnetic field has independently found its description in a fundamental continuity relation \cite{lipkin1964existence} for the flow of optical chirality \cite{PhysRevLett.104.163901}, derivable solely from the electromagnetic governing equations.

Apart from their disconnected origins, multiple issues are known to complicate matter and optical chirality measures, e.g.: (1) the requirement for a preferred choice of reference axis or structure \cite{PhysRevLett.133.268001, tao2026chiral, zabrodsky1995continuous}, (2) the existence of false chiral zeros \cite{vavilin2022multidimensional,moudgal2025multiscale,weinberg1997chiral}, and (3) the transferrable utility and/or lack of proposed measurements to assess their values \cite{BudaQuantifying,petitjean2003chirality,cha2024graph}. In this Letter, we resolve these difficulties through the introduction of a unified complex-valued pseudoscalar metric that is derived directly from the coupled governing equations for light and matter. Our approach bridges the length and energy scales from molecular to nanophotonic domains, and is here applied to a family of Born-Kuhn structures  \cite{Born1918,Kuhn1930}  composed of identical plasmonic nanorods \cite{yin2013interpreting,bourgeois2022polarization}.

The continuity equation for optical chirality density \cite{lipkin1964existence,PhysRevLett.104.163901} provides a cornerstone for the quantification of a material object's excitational chirality in the absence of externally applied driving fields \cite{8f2s-rjgy}. Specifically, in the low-loss limit, the time- and volume-averaged optical chirality radiated from the object's $\Gamma$th eigenmode $\langle\Phi\rangle_\Gamma \equiv\int d{\bf x}\nabla\cdot\langle\boldsymbol{\varphi}\rangle_\Gamma=\int d{\bf x}\langle\mathcal{J}\rangle_\Gamma$ is a pseudoscalar reporter of its intrinsic excitational chirality through global characterization of its emitted eigenfield polarization states \cite{8f2s-rjgy}, \textcolor{black}{where $\boldsymbol{\varphi}=(c/8\pi)[{\bf E}\times(\nabla\times{\bf B})-{\bf B}\times(\nabla\times{\bf E})]$ and $\mathcal{J}=-(1/2)[{\bf E}\cdot\nabla\times{\bf J}+{\bf J}\cdot\nabla\times{\bf E}]$ are the flux and source of optical chirality, respectively \cite{PhysRevLett.104.163901}.} Explicitly, $\langle\Phi\rangle_\Gamma$ can be interpreted as the angle-integrated $S_3$ Stokes parameter of the radiated fields \cite{poulikakos2016optical,bezard2026absolute,PhysRevA.98.013837}, since $\langle\Phi\rangle_\Gamma =({\omega^2_\Gamma}/{2})\int r^2d\Omega\,\hat{\bf n}\cdot\langle \mathbf{s}\rangle_\Gamma,$ where $\langle \mathbf{s}\rangle_\Gamma$ is the canonical optical spin angular momentum density \cite{barnett2010rotation, vernon_decomposition_2024, 8f2s-rjgy,PhysRevA.83.021803,PhysRevA.85.063810}. Therefore, $\langle\Phi\rangle_\Gamma$ quantifies the chirality of each eigenmode of frequency $\omega_\Gamma=k_\Gamma c$ through the dissymmetry in local circular polarization content of the associated radiation fields.

\begin{figure}
    \centering
    \includegraphics[width=\linewidth]{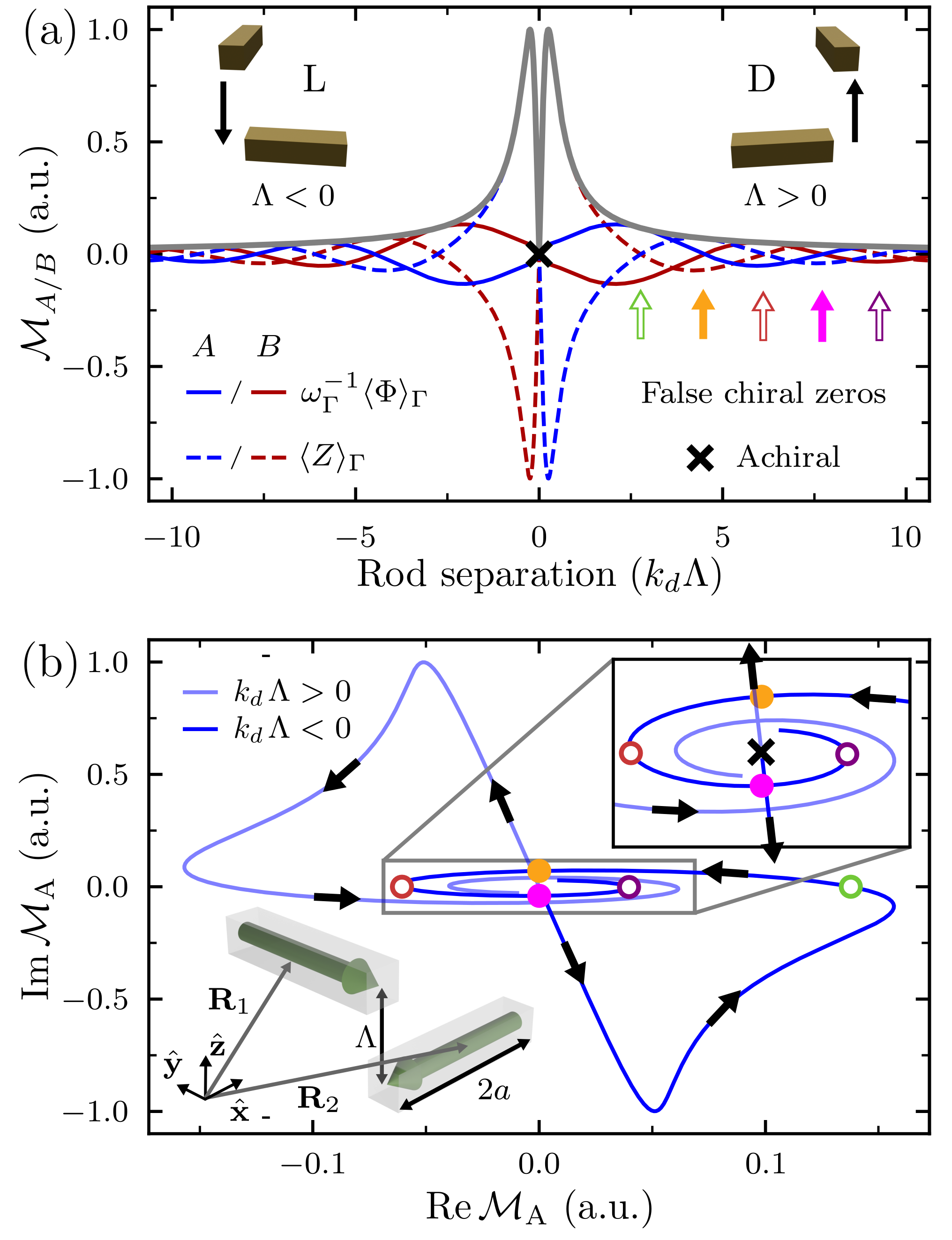}
    \caption{Composite light-matter eigenmode chirality of the BK$_2$ system under continuous deformation in $k_d\Lambda$. Blue and red solid (dashed) lines in (a) are the radiated optical chirality $\langle\Phi\rangle_\Gamma=-\omega_\Gamma\text{Im}\{ \mathcal{M}_\Gamma \}$ (matter chirality $\langle Z\rangle_\Gamma=\text{Re}\{ \mathcal{M}_\Gamma \}$) of the $\Gamma\in\{A,B\}$ eigenmodes, respectively, while the gray line is $|\mathcal{M}_\Gamma|$. FCZs are indicated for $k_d\Lambda>0$ by the colored arrows. Corresponding points are plotted in the complex plane in (b) for $\mathcal{M}_A(k_d\Lambda)$. The zoom-in panel shows $\mathcal{M}_A$ passing through a true chiral zero at $k_d\Lambda = 0$, indicated by $\times$. All axes are normalized as $\mathcal{M}_\Gamma/\textrm{max}|\mathcal{M}_\Gamma|$.}
    \label{figure1}
    %, where the green point corresponds to an antibonding eigenmode of the D enantiomer (inset)
\end{figure}
In the case of the Born-Kuhn dimer system (BK$_2$) shown in Fig. \ref{figure1}, the optical
chirality radiated from the bonding ($\Gamma = B$) and antibonding ($\Gamma = A$) dipolar eigenmodes
$\bm{\xi}_{\Gamma}=[{\bf d}_1^\Gamma,{\bf d}_2^\Gamma]=[\hat{\bf x}{d}_1^\Gamma,\hat{\bf y}{d}_2^\Gamma]$ is \cite{8f2s-rjgy}
\begin{equation}
%\begin{split}
   \langle \Phi \rangle_\Gamma = -c k_\Gamma^5 \frac{\Lambda}{r_{12}}j_1(k_\Gamma r_{12})\text{Re}\big[d^\Gamma_1(d^\Gamma_2)^*\big],
%   &=
%-\frac{ck_\Gamma^5}{2r_{12}}\Big(\frac{e}{m\omega_\Gamma}\Big)^2j_1(k_\Gamma r_{12})
%\Big({\mathbf L}_{12}^{\,*}\cdot{\mathbf p}_2+{\mathbf L}_{21}^{\,*}\cdot{\mathbf p}_1\Big).
\label{phi_scaig}
%\end{split}
\end{equation}
where $j_1$ is the spherical Bessel function of the first kind. The separation distance $r_{12}=|{\bf R}_1-{\bf R}_2|=\sqrt{\Lambda^2+2a^2}$, where $\Lambda = z_1 - z_2$, and ${\bf R}_1=\hat{\bf x}a+\hat{\bf z}z_1$ and ${\bf R}_2=\hat{\bf y}a+\hat{\bf z}z_2$ are the nanorod center positions. Fig. \ref{figure1}(a) displays the behavior of $\langle \Phi\rangle_\Gamma$ for each eigenmode (red, blue solid lines) as $k_d\Lambda$ is varied across near-, intermediate-, and far-field regimes with $\omega_d=k_dc$ being the longitudinal dipole resonance frequency. When the two nanorods are separated by distances that are well within the plasmon resonance wavelength, i.e., $|k_d \Lambda| \ll 1$, $\langle \Phi\rangle_A$ and $\langle \Phi\rangle_B$ carry opposite signs that flip as $k_d \Lambda$ passes through zero, reflecting the change in enantiomeric state of the underlying static structure and its eigenexcitations, passing through an achiral configuration at $k_d \Lambda =0$ where $\langle \Phi\rangle_\Gamma=0$. Despite this bisignate behavior around $k_d \Lambda =0$, an issue arises at the zeros of $j_1(k_\Gamma r_{12})$. Even though the underlying BK$_2$ structure remains geometrically chiral and in the same enantiomeric state with truly chiral eigenmodes \cite{barron1986true}, $\langle \Phi\rangle_\Gamma=0$. Such zero crossings (denoted by filled colored vertical arrows for $k_d \Lambda > 0$) are examples of ‘false chiral zeros' (FCZs) \cite{vavilin2022multidimensional,moudgal2025multiscale,weinberg1997chiral}, where $\langle \Phi\rangle_\Gamma$ fails to assign a hand despite the material structure and its eigenmodes exhibiting a definite handedness.

FCZs have been studied in other material systems \cite{PhysRevB.110.174112,banik2016orientation,moudgal2025multiscale} and have been found to be ubiquitous, where chiral objects with sufficient degrees of freedom possess the property of \emph{chiral connectedness} \cite{buda1992hausdorff,mislow1993shape,weinberg1997chiral,PhysRevX.4.011003,RevModPhys.71.1745}, allowing for continuous deformations between enantiomeric partners without passing through an achiral configuration. As a result, FCZs are generally endemic to all bisignate, real-valued scalar chirality metrics, motivating their extension to more general measures that lie beyond the real number system \cite{vavilin2022multidimensional}.

Beyond the fact that $\langle \Phi\rangle_\Gamma\in{\mathbb R}$ and is susceptible to FCZs, reporting only on the chirality stored in the radiation field, $\langle \Phi\rangle_\Gamma$ does not characterize how chirality is recorded or dissipated in the sourcing matter. Motivated by these deficiencies, we derive a continuity relation for composite light-matter chirality flow starting from the optical chirality transfer rate ${\cal P}=\int d{\bf x}\mathcal{J}$ in analogy to the continuity of energy flow between a single dipole oscillator ${\bf d}=-e{\bf r}$ of effective mass $m$ \cite{Cherqui_2014} and the electromagnetic field (see End Matter), where $\mathcal{J}=-(1/2)[{\bf E}\cdot\nabla\times{\bf J}+{\bf J}\cdot\nabla\times{\bf E}]$ is expressed in terms of the material current density $\bf J$ and electric field $\bf E$. Elimination of $\bf J$ from ${\cal P}$ using Maxwell's equations results in the known continuity equation for optical chirality \cite{PhysRevLett.104.163901, 8f2s-rjgy}.

Oppositely, eliminating $\bf E$ from ${\cal P}$ using the dipole's Newton equations (Eqs. \eqref{newtoneqs}) produces the continuity equation
\begin{equation}
-\frac{d}{dt}\Big[m\dot{\bf r}\cdot\nabla_{\bf R}\times\dot{\bf r}\Big]+m\ddot{\bf r}\cdot\nabla_{\bf R}\times\dot{\bf r}={\cal P},
\label{CE_Z}
\end{equation}
implicitly identifying a matter chirality $Z=m\dot{\bf r}\cdot\nabla_{\bf R}\times\dot{\bf r}\in{\mathbb R}$. Here, we restrict the center of mass motion to the $\dot{\bf R}={\bf 0}$ limit appropriate for nanophotonic objects immobilized on a supporting substrate, neglecting contributions to $Z$ that would emerge in other systems, such as from the nonadiabatic nuclear and electronic couplings present in certain molecules \cite{tao2026chiral} or the connections between mechanical motion and the magnetic field characteristic of the Einstein–de Haas effect \cite{einstein1915experimenteller,peng2026phase}. 
\textcolor{black}{In the $\dot{\bf R}={\bf 0}$ limit (Eq. \eqref{CE_Z}), the composite  continuity equation expressing the conservation of total light-matter chirality becomes
\begin{equation}
\frac{d}{dt}\Big[Z+\int d{\bf x}\,{C}\Big]=m\ddot{\bf r}\cdot\nabla_{\bf R}\times\dot{\bf r}-\int r^2d\Omega\,\hat{\bf n}\cdot\boldsymbol{\varphi}
\end{equation}
in direct analogy to the conservation of total energy (Eq. \eqref{energy}). Here, the first term on the right hand side is a chiral force acting to dissipate chirality in the matter, while the second term expresses the radiation of optical chirality to infinity due to radiatively open boundary conditions.}

The matter chirality derived for a single dipole oscillator under driven conditions may be generalized to many electromagnetically interacting dipoles and expressed in the absence of external driving forces. By noting that the kinetic momentum associated with the $i$th oscillator coordinate ${\bf r}_i$ is $m\dot{\bf r}_i={\bf p}_i+(e/c){\bf A}({\bf R}_i),$ it is evident that $\nabla_{{\bf R}_i}\times\dot{\bf r}_i=(e/mc){\bf B}({\bf R}_i),$ where ${\bf B}({\bf R}_i)=\nabla_{{\bf R}_i}\times{\bf A}({\bf R}_i)$ is the  magnetic field from all other dipoles evaluated at ${\bf R}_i$. Thus, the eigenmode material chirality for a collection of interacting dipoles is defined as
\begin{equation}
%Z_\Gamma=-\frac{1}{c}\sum_{i\neq j}{\bf j}^{\Gamma}_i\cdot{\bf B}^{\Gamma}_j({\bf R}_i),\\
Z_\Gamma=-\frac{1}{c}\int d{\bf x}\,{\bf J}_{\Gamma}({\bf x})\cdot{\bf B}_{\Gamma}({\bf x})
\label{MC}
\end{equation}
where ${\bf J}_{\Gamma}({\bf x})=\sum_i(-e\dot{\bf r}^{\Gamma}_i)\delta({\bf x}-{\bf R}_i)=\sum_i{\bf j}^{\Gamma}_i\delta({\bf x}-{\bf R}_i)$ is the current density of the $i$th oscillator within the $\Gamma$th eigenmode and \textcolor{black}{${\bf B}_{\Gamma}({\bf x})=\sum_{i=1,2}{\bf g}({\bf x},{\bf R}_i)\times{\bf d}^\Gamma_i$ is the magnetic field excluding self interactions expressed in terms of the magnetic Green's function ${\bf g}({\bf x},{\bf R}_i)$ \cite{Jackson2003} and electric dipole moments ${\bf d}^\Gamma_i$.} Physically, $Z_\Gamma$ is a gauge invariant quantity characterizing the manner in which magnetic flux from the distinct $j \ne i$ oscillators braid the current loop of each oscillator $i$ within eigenmode $\Gamma$ \cite{berger1984topological}.

For the BK$_2$ system, $Z_\Gamma=-({1}/{c})[{\bf j}^{\Gamma}_1\cdot{\bf B}^{\Gamma}_2({\bf R}_1)+{\bf j}^{\Gamma}_2\cdot{\bf B}^{\Gamma}_1({\bf R}_2)].$ Upon time averaging over complex fields, $\langle Z\rangle_\Gamma=-({1}/{2c})\textrm{Re}\Big[{\bf j}^{\Gamma}_1\cdot{\bf B}^{\Gamma}_2({\bf R}_1)^*+{\bf j}^{\Gamma}_2\cdot{\bf B}^{\Gamma}_1({\bf R}_2)^*\Big]$, which takes the form
\begin{align}
\langle Z\rangle_\Gamma
=
k_\Gamma^4 \frac{\Lambda}{r_{12}}y_1(k_\Gamma r_{12})\text{Re}\big[d^\Gamma_1(d^\Gamma_2)^*\big],
%&=
\end{align}
where $y_1$ is the irregular spherical Bessel function. Notably, $\langle Z\rangle_A\neq0$ and $\langle Z\rangle_B\neq0$ for all $\Lambda\neq0$, except at the FCZs denoted by open colored vertical arrows in Fig. \ref{figure1}(a), reflecting the fact that the BK$_2$ eigenmodes are truly chiral for all nonzero nanorod separations \cite{8f2s-rjgy}. By comparison with the radiated optical chirality $\langle\Phi\rangle_\Gamma =\int d{\bf x}\langle\mathcal{J}\rangle_\Gamma=-(\omega_\Gamma/2c)\int d{\bf x}{\textrm{Im}}[{\bf J}_\Gamma\cdot{\bf B}^*_\Gamma]$, the matter chirality can be expressed similarly as $\langle Z\rangle_\Gamma=-(1/2c)\int d{\bf x}{\textrm{Re}}[{\bf J}_\Gamma\cdot{\bf B}^*_\Gamma].$ Taken together, the pair form the composite measure
\begin{equation}
\begin{split}
 {\mathcal{M}}_\Gamma 
 &=\langle Z\rangle_\Gamma-\frac{i}{\omega_\Gamma}\langle{\Phi}\rangle_\Gamma\\
 %&= \textcolor{black}{\frac{ik}{2c}\sum'_{ij}{\bf j}_i^\Gamma\cdot\left[{\bf g}({\bf R}_i,{\bf R}_j)\times{\bf p}_j^\Gamma\right]^*}\\
 &= \textcolor{black}{-\frac{1}{2c}\left[{\bf j}_1^\Gamma\cdot\left({\bf g}_{12}\times{\bf d}_2^\Gamma\right)^*+{\bf j}_2^\Gamma\cdot\left({\bf g}_{21}\times{\bf d}_1^\Gamma\right)^*\right]}\\
 %&=\textcolor{black}{\frac{ik_\Gamma}{2c}\left\{{\bf j}_1^\Gamma\cdot\left[{\bf g}_{12}\times{\bf p}_2^\Gamma\right]^*+[1\leftrightarrow2]\right\}}\\
 &=ik_\Gamma^4\frac{\Lambda}{r_{12}}h^{(2)}_1(k_\Gamma r_{12})\text{Re}\big[d^\Gamma_1(d^\Gamma_2)^*\big]
% &=\sum_{i\neq j}
%\frac{k^4_\Gamma}{2ir_{12}}\Big(\frac{e}{m\omega_\Gamma}\Big)^2h^{(2)}_1(k_\Gamma r_{ij})
%\left(
%{\mathbf L}_{ij}^*\cdot{\mathbf p}_j+{\mathbf L}_{ji}^*\cdot{\mathbf p}_i\right) 
%&=\sum_{i\neq j}
%\frac{i{\omega_\Gamma}\Big(\frac{ek_\Gamma r_{ij}}{m}\Big)^2
%\left(
%{\mathbf L}_{ij}^*\cdot{\mathbf p}_j
%+{\mathbf L}_{ji}^*\cdot{\mathbf p}_i\right) h^{(2)}_1(k_\Gamma r_{ij})
    \label{MPS}
    \end{split}
\end{equation}
characterizing the light-matter chirality of the $\Gamma$th eigenmode, where \textcolor{black}{${\bf g}_{12}\equiv{\bf g}({\bf R}_1,{\bf R}_2) = -ik(ik/r_{12}-1/r_{12}^2)({\bf R}_1-{\bf R}_2)e^{ikr_{12}}/{r_{12}}=-k^3({\bf R}_1-{\bf R}_2)h^{(1)}_1(kr_{12})/r_{12}$ and $h^{(1)}_1$ and $h^{(2)}_1$ are the spherical Hankel functions of the first and second kind}. Notably, ${\cal M}_\Gamma\in{\mathbb C}$ is a unified pseudoscalar measure of intrinsic eigenmode chirality that is derived directly from the governing dynamical equations for matter and field\footnote{Certain classes of chiral objects exist with all dipole moments oriented along a common (arbitrary) $\hat{\bf e}$ \cite{vavilin2022multidimensional} that do not produce elliptically polarized radiation in any direction, implying $\langle \Phi \rangle_\Gamma =0$. The magnetic flux loops of the constituent dipoles in these objects do not cross such that $\langle Z \rangle_\Gamma=0$.  Nevertheless, the handedness of such objects may be imprinted upon their near-field Stokes vectors and/or far-field wavefront phases \cite{machfuudzoh202}.}, and provides a rigorous foundation for the family of momentum pseudoscalar metrics (see End Matter) \cite{PhysRevLett.133.268001, 8f2s-rjgy,tao2026chiral}.

By characterizing eigenmode chirality from the broader position offered by ${\cal M}_\Gamma,$ Fig. \ref{figure1}(a) additionally displays $\langle Z\rangle_\Gamma={\textrm{Re}}\{{\cal M}_\Gamma\}$ (red, blue dashed lines) and the complex norm $|{\cal M}_\Gamma|$ (gray solid line) versus separation $k_d\Lambda$ for the BK$_2$ system. At the locations of the FCZs of $\langle \Phi\rangle_\Gamma=-\omega_\Gamma{\textrm{Im}}\{{\cal M}_\Gamma\}$ (denoted by the vertical filled arrows, where $\langle \Phi\rangle_\Gamma=0$), it is evident that $\langle Z\rangle_\Gamma\neq0.$ Thus, $\langle Z\rangle_\Gamma$ provides a hand to the $\Gamma$th eigenmode when $\langle \Phi\rangle_\Gamma$ fails to do so. Oppositely, at those separations $k_d\Lambda$ where $\langle Z\rangle_\Gamma=0,$ indicated by the empty vertical arrows in Fig. \ref{figure1}a, $\langle \Phi\rangle_\Gamma\neq0$ and, again, a hand is provided. It is only for the achiral BK$_2$ geometry defined by $k_d\Lambda=0$, where $\langle Z\rangle_\Gamma=\langle \Phi\rangle_\Gamma=0$. Thus, the issue of FCZs is alleviated by consideration of the composite light-matter chirality ${\cal M}_\Gamma.$

\begin{figure}
    \centering
\includegraphics[width=\linewidth]{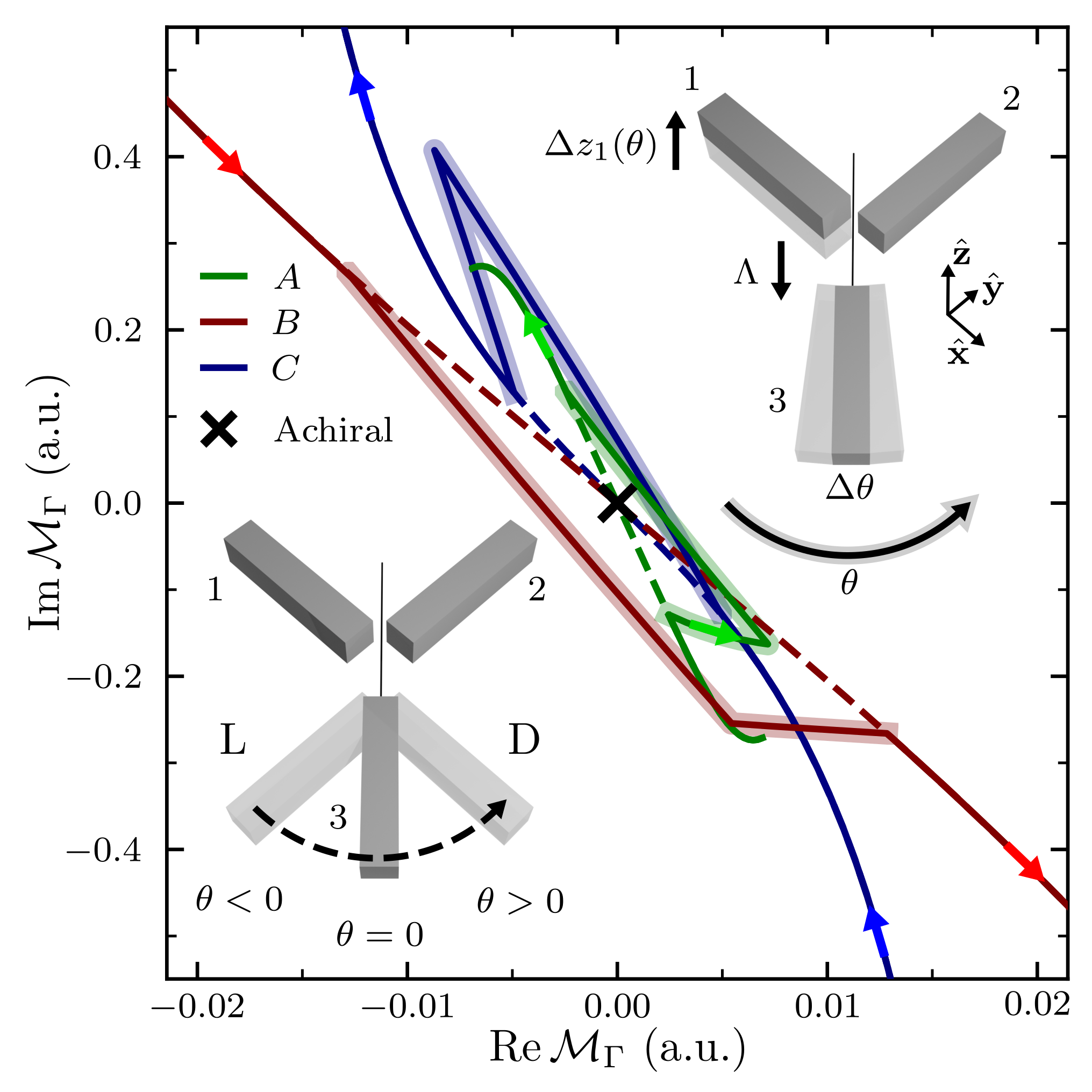}
    \caption{Eigenmode chiralities $\mathcal{M}_\Gamma$ of the chirally connected BK$_3$ system with $\Gamma\in\{A,B,C\}$ under continuous structural deformation between two enantiomers at fixed $k_d \Lambda =-0.15$. Dashed traces correspond to $(\theta,\delta z_1 =0)$-trajectories that pass through the achiral geometry at $\theta=0$ (denoted by $\boldsymbol{\times}$), while solid traces correspond to $(\theta,\delta z_1 \ne0)$-trajectories that detour around the achiral geometry passing entirely through chiral configurations. All axes are normalized as $\mathcal{M}_\Gamma/\textrm{max}
    _\Gamma|\mathcal{M}_{\Gamma}|$.}
    \label{figure2}
\end{figure}
For BK$_2$ systems with fixed geometric handedness as $|k_d \Lambda|$ increases, we interpret the oscillation in sign of ${\textrm{Re}}\{\mathcal{M}_\Gamma\}$ and ${\textrm{Im}}\{\mathcal{M}_\Gamma\}$ as distinct analogs of the wagon wheel effect \cite{purves1996wagon}, here arising from the discord in length scales between the inter-rod separation $\Lambda$ and plasmon resonance wavelength $2\pi c/\omega_\Gamma$. These examples of chiral aliasing are anticipated to be ubiquitous, perhaps helping to explain the general existence of eigenmodes of opposite handedness in different spectral regions (e.g., vibrational and electronic states in molecular systems arising from distinct Hamiltonians within the Born-Oppenheimer approximation \cite{tao2026chiral}) on top of a fixed geometric scaffold  \cite{kumar2023photonically, PhysRevLett.133.268001}. Fig. \ref{figure1}(b) displays ${\textrm{Re}}\{{\cal M}_A\}$ and ${\textrm{Im}}\{{\cal M}_A\}$ parametrized in increasing $|k_d\Lambda|,$ beginning at $k_d\Lambda=0$ (indicated by $\times$) and progressing for inter-rod separations spanning into the far-field. The colored markers correspond to the same $k_d\Lambda$ separation values of the FCZs indicated by the colored arrows in Fig. \ref{figure1}(a).

Fig. \ref{figure2} examines a chiral BK trimer system (BK$_3$; see top inset) that exhibits a richer variety of FCZs, together with enantiomeric states that are chirally connected. Two geometric deformation pathways for ${\cal M}_\Gamma$ are considered where enantiomeric partners are (1) connected by rotation of nanorod 3 through the angle $\theta\in[-\pi/4,\pi/4]$, passing through an achiral geometry at $\theta=0$ (denoted by $\boldsymbol{\times}$ where ${\textrm{Re}}\{{\cal M}_\Gamma\}={\textrm{Im}}\{{\cal M}_\Gamma\}=0$); or (2) chirally connected in the region $\theta\in[-\Delta\theta,\Delta\theta]$ through the deformation pathway parametrized by $\Delta z_1(\theta) =  z_{\textrm{max}}(1-|\theta|/\Delta\theta)$, where $\Delta\theta=\pi/12$, passing entirely through chiral geometries. Solid (dashed) curves trace the (a)chiral paths for each of the $\Gamma\in\{A,B,C\}$ eigenmodes. Thus, the existence of a path that lacks a synchronous zero crossing of ${\textrm{Re}}\{{\cal M}_\Gamma\}$ and ${\textrm{Im}}\{{\cal M}_\Gamma\}$ is a  distinguishing feature of chirally connected objects.

While ${\cal M}_\Gamma$ quantifies intrinsic chirality in the absence of external forces, experimental characterization of chiral objects requires use of a probe. In the presence of an external optical driving field, extinction cross sections for optical $\igmas^{\textrm{opt}}_{\textrm{ext}}(\omega)$ and matter $\igmas^{\textrm{mat}}_{\textrm{ext}}(\omega)$ chirality can be defined in analogy to those for energy (see End Matter) \cite{8f2s-rjgy}. Due to the causal nature of their common governing equations (Eq. \eqref{newtoneqs}), $\igmas^{\textrm{opt}}_{\textrm{ext}}(\omega)$ and $\igmas^{\textrm{mat}}_{\textrm{ext}}(\omega)$ are analytically connected Kramers-Kronig partners \cite{toll1956causality} with $\igmas_{\textrm{ext}}(\omega)=\igmas^{\textrm{mat}}_{\textrm{ext}}(\omega)+i\igmas_{\textrm{ext}}^{\textrm{opt}}(\omega)$, such that measurement of one spectrum determines the other. Self-referencing schemes measuring the amplitude, polarization, and phase of forward-scattered light in principle offer direct access to both components via direct measurement \cite{nayak2026decoding, crdw-pxcs}.

In the case of the BK$_2$ structure illuminated by a $\phi=\pi/4$ linearly polarized plane-wave field, the complex-valued matter-field chirality cross section for extinction is
%\begin{widetext}
\begin{equation}
\begin{aligned}
   {\igmas}_{\textrm{ext}}(\omega)
&= \frac{2\pi\omega}{c} \Big[ i  \bigg( \frac{\alpha_2 -\alpha_1}{1 - g^2 \alpha_1 \alpha_2}\bigg)  \\
&\ \ \ \ \ \ \ \ \ \ \ -   \frac{2g\alpha_1\alpha_2 }{1 - g^2 \alpha_1 \alpha_2}\sin(k\Lambda) \Big],
\label{chiext}
\end{aligned}
\end{equation}
%\end{widetext}
where the two dipoles ($i=1,2$) have the natural frequencies $\omega_i$, total decay rates $\gamma_i,$ effective masses $m_i$, and polarizabilities $\alpha_i(\omega)=(e^2/m_i)/(\omega_i^2-\omega^2-i\gamma_i\omega)$, and are coupled with dipole-dipole coupling strength $g(\omega)= \hat{\bf{R}}_1 \cdot \tensor{\bf{G}}(\omega) \cdot \hat{\bf{R}}_2$ with vacuum dipole relay tensor $\tensor{\bf{G}}({\bf R}_1,{\bf R}_2,\omega)=[(\omega/c)^2+\nabla_{{\bf R}_1}\nabla_{{\bf R}_2}]e^{i(\omega/c)|{\bf R}_1-{\bf R}_2|}/|{\bf R}_1-{\bf R}_2|$ \cite{Jackson2003}. The pole structure inherent to Eq. \eqref{chiext} provides insight into strategies to extremize the chiral response of the BK$_2$ structure through modification of its material parameters.

\begin{figure}
    \centering
\includegraphics[width=3.25in]{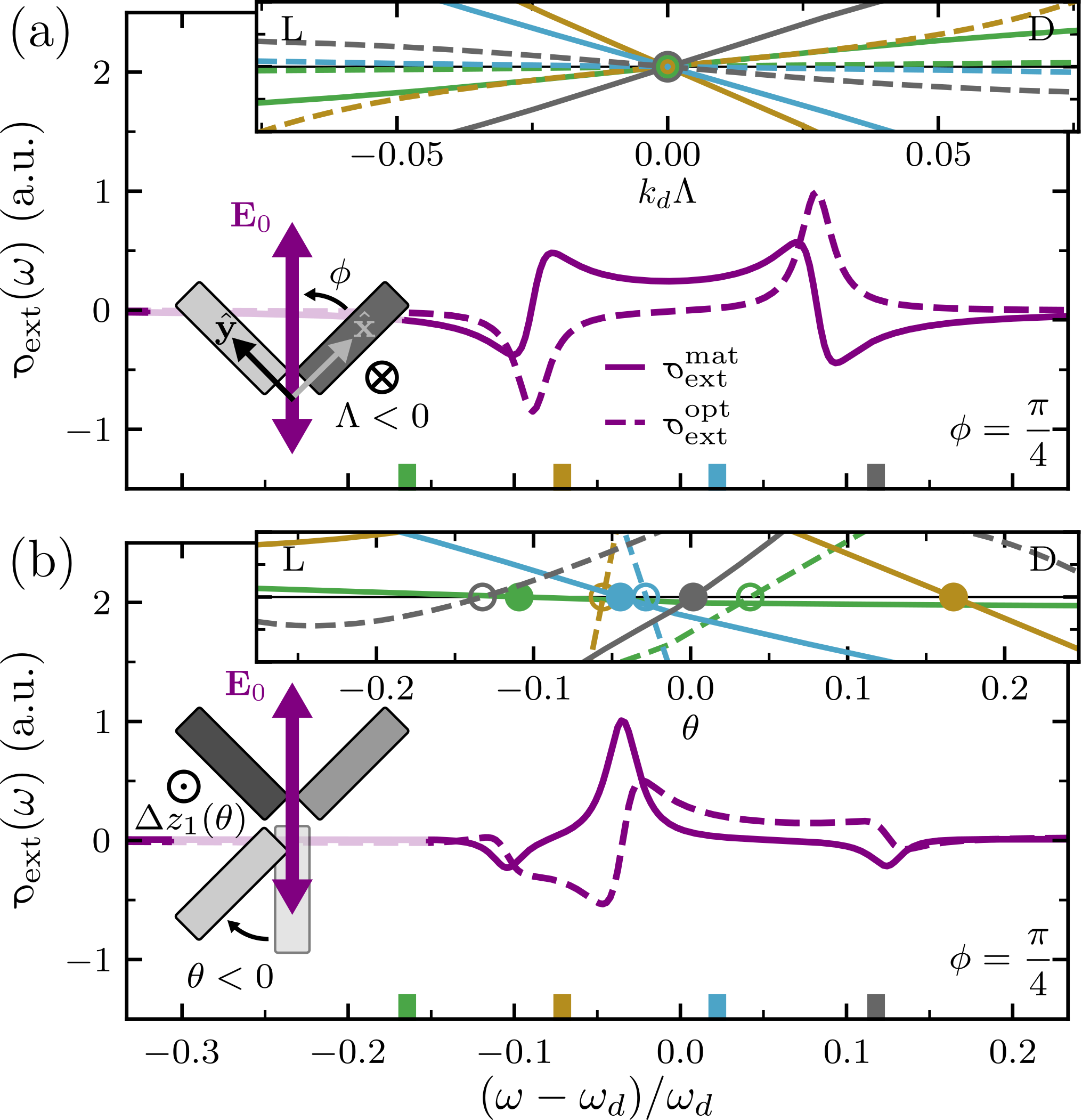}
    \caption{Chiral extinction cross section spectra $\igmas_{\textrm{ext}}(\omega)$ for the L enantiomers of the (a) BK$_2$ ($k_d \Lambda =-0.08$) and (b) BK$_3$ ($\theta =-\pi/4$) structures under $\phi=\pi/4$ linearly polarized optical plane wave excitation ${\bf E}_0$, with $\igmas_{\textrm{ext}}^{\textrm{mat}}(\omega) = \text{Re}\{\igmas_{\textrm{ext}}(\omega)\}$ (solid purple) and $\igmas_{\textrm{ext}}^{\textrm{opt}}(\omega) = \text{Im}\{\igmas_{\textrm{ext}}(\omega)\}$ (dashed purple). Continuous deformation of $\igmas_{\textrm{ext}}^{\textrm{mat}}(\omega)$ (solid traces) and $\igmas_{\textrm{ext}}^{\textrm{opt}}(\omega)$ (dashed traces) at various probe frequencies (colored markers) is displayed in panel (a) inset versus $k_d\Lambda$, passing through an achiral structure at $k_d\Lambda = 0,$ or panel (b) inset versus $\theta,$ avoiding the achiral structure and never crossing zero simultaneously, as indicated by the filled and open circles.}
    \label{figure3}
\end{figure}
Fig. \ref{figure3} displays the real and imaginary components of the light-matter extinction cross section spectrum $\igmas_{\textrm{ext}}(\omega)$ for the L enantiomers of the (a) BK$_2$ and (b) BK$_3$ structures with $\igmas_{\textrm{ext}}^{\textrm{mat}}(\omega)$ (purple traces) and $\igmas_{\textrm{ext}}^{\textrm{opt}}(\omega)$ (dashed purple traces) under linearly polarized optical plane wave excitation with polarization ${\bf E}_0$ at angle $\phi = \pi/4$. The insets display the evolution of $\igmas_{\textrm{ext}}^{\textrm{mat}}(\omega)$ (solid traces) and $\igmas_{\textrm{ext}}^{\textrm{opt}}(\omega)$ (dashed traces) at various probe frequencies (indicated by colored markers) for the L to D transformation of the (a) BK$_2$ structure deformed in $k_d\Lambda$, passing through an achiral configuration at $\Lambda=0,$ or for the (b) chirally connected BK$_3$ structure deformed in $\Delta z_1(\theta)$, passing entirely through chiral configurations. Whereas all of the traces in the panel (a) inset cross through zero at $k_d\Lambda=0$ (i.e., at the true chiral zero), the values of $\theta$ where the real and imaginary components of $\igmas_{\textrm{ext}}(\omega, \theta)$ cross zero in the panel (b) inset, marked as filled and empty circles, respectively, never coincide. This behavior of the $\igmas_{\textrm{ext}}(\omega)$ observable highlights its ability to identify and characterize chirally connected objects.

The constraints imposed by the inseparability between matter and field give rise to fundamental connections that relate the structural chirality of a material object to the excitational chirality expressed by its intrinsic excitations and associated electromagnetic fields. However, metrics established to assess chirality are limited to perspectives that consider matter and field in isolation and bear related complications that hinder their general applicability. Here, we propose a unified chirality metric that is derived from the coupled dynamical equations governing the interactions between light and matter and is directly measurable via optical polarimetry. While transferable across the energy and length scales spanning molecular \cite{tang2011enhanced, kondru1998atomic}, solid-state crystalline \cite{Ishito2023,PhysRevLett.134.026902}, and nanophotonic objects \cite{Lee2018,cho2023bioinspired}, and their combinations, we demonstrate resolution of these complications through application to a generic set of structurally chiral nanophotonic objects, including those that exhibit FCZs and are chirally connected. Explicit calculations of the proposed matter-field chirality extinction cross section are presented under linearly-polarized plane wave illumination to spectroscopically distinguish between chirally connected and achiral pathways. Further extension to incorporate near-field probes could offer new insights into the relationship between material and electromagnetic chirality at sub-diffraction-limited length scales, potentially including the chirality-induced spin selectivity (CISS) effect \cite{bloom2024chiral}. \textcolor{black}{Application of ${\cal M}_\Gamma$ to periodic crystalline materials extends the toolkit for quantification and identification of chiral phonons \cite{juraschek2025chiral} and their properties, including asymmetric transport \cite{chen2022chiral} and handed interactions with light \cite{PhysRevLett.134.026902}. Nanophotonic chiral sensing platforms \cite{warning2021nanophotonic,both2022nanophotonic} represent yet another frontier benefiting from rigorous measures of light-matter chirality, where the rational design of materials exhibiting maximal chiral responses is of key interest.}

%%%%%%%%%%%%%%%%%%%%%%%%%%%%%%%%%%%%%%%%%%%%%%%%%
\begin{acknowledgments}
Theoretical development of the proposed complex-valued matter-field chirality metric ${\cal M}_\Gamma$ was supported by the National Science Foundation under award No. CHE-2516408. Spectroscopic interpretation and measurement of the generalized chirality cross section $\igmas_{F}(\omega)$ was supported by the National Science Foundation through the Center for Single-Entity Nanochemistry and Nanocrystal Design, part of the Centers for Chemical Innovation program (CHE-2503933). 
\end{acknowledgments}
%%%%%%%%%%%%%%%%%%%%%%%%%%%%%%%%%%%%%%%%%%%%%%%%%

\bibliography{references}

% --- END MATTER (PLACED AFTER REFERENCES) ---

\appendix

\newpage

\subsubsection{Optical chirality}
Optical chirality density $C = (8\pi)^{-1}[ {\bf E} \cdot  {\nabla} \times {\bf E} + {\bf B} \cdot {\nabla} \times{\bf B}]$ is conserved under Maxwell flow according to
\begin{equation}
\dot{C}+\nabla\cdot\boldsymbol{\varphi}=\mathcal{J},
\label{Cflow}
\end{equation}
where $\boldsymbol{\varphi}=(c/8\pi)[{\bf E}\times(\nabla\times{\bf B})-{\bf B}\times(\nabla\times{\bf E})]$ and $\mathcal{J}=-(1/2)[{\bf E}\cdot\nabla\times{\bf J}+{\bf J}\cdot\nabla\times{\bf E}]$ are the flux and source of optical chirality \cite{PhysRevLett.104.163901}. In the absence of external fields, Eq. \eqref{Cflow} is also satisfied for the eigenfields associated with a sourcing object's eigenmode excitations ${\bm\xi}_\Gamma$. Assuming that all quantities vary in time as $X(t)=Xe^{-i\omega_\Gamma t}$ where \textcolor{black}{$|\textrm{Im}\{\omega_\Gamma\}|\ll|\textrm{Re}\{\omega_\Gamma\}|$}, the time- and volume-averaged rate $\langle(d/dt){C}\rangle_\Gamma\approx0$ for each mode $\Gamma$, which defines the radiated flux of optical chirality,
\begin{equation}
%\begin{split}
\langle\Phi\rangle_\Gamma \equiv\int d{\bf x}\nabla\cdot\langle\boldsymbol{\varphi}\rangle_\Gamma=\int d{\bf x}\langle\mathcal{J}\rangle_\Gamma.
\label{lowloss}
%\end{split}
\end{equation}
Partial integration together with Faraday's law connect the following representations 
\begin{equation}
\begin{split}
\langle\Phi\rangle_\Gamma 
&=-\frac{1}{2}\int d{\bf x}{\textrm{Re}[{\bf J}_\Gamma\cdot\nabla\times{\bf E}_\Gamma^*]}\\
&=-\frac{\omega_\Gamma}{2c}\int d{\bf x}{\textrm{Im}}[{\bf J}_\Gamma\cdot{\bf B}^*_\Gamma].
\end{split}
\end{equation}
\textcolor{black}{In general, the cycle-averaged optical chirality density $\langle C\rangle$, flux $\langle\boldsymbol{\varphi}\rangle$, and source $\langle\mathcal{J}\rangle$ decay as $e^{2\text{Im}\{\omega_\Gamma\}t}$, such that
\begin{equation}
    2\textrm{Im}\{\omega_\Gamma\}\int d{\bf x} \langle C \rangle_\Gamma+\langle\Phi\rangle_\Gamma = \int d{\bf x} \langle\mathcal{J}\rangle_\Gamma,
\end{equation} which reduces to Eq. \eqref{lowloss} in the low-loss limit or under steady-state driving. Eq. \eqref{lowloss} is also valid whenever $\int d{\bf x} \langle C \rangle_\Gamma = 0$.} Here and throughout, $\langle \mathcal{A} \rangle_\Gamma$ indicates time averaging of the quantity $\mathcal{A}$ over the oscillatory period $2\pi/\omega_\Gamma$ of eigenmode $\Gamma$. All volume integrations extend into the far-field and contain the sourcing material structure within their bounding surface with surface normal $\hat{\bf n}$.

%, including the material Hamiltonian without its a priori knowledge

\subsubsection{Energy conservation}
The matter-field chirality conservation relation is derived in analogy to the continuity of composite matter-field energy. For simplicity, we consider a system composed of a single electric dipole oscillator ${\bf d}=-e{\bf r}$ of natural frequency $\omega_0$ coupled to the electromagnetic field with energy transfer rate $P=\int d{\bf x}{\bf E}\cdot{\bf J}.$ The system's dynamics are governed by the Newton equations
\begin{equation}
    \begin{aligned}
m\ddot{\bf r}&=-m\omega_0^2{\bf r}-e\Big({\bf E}+\frac{1}{c}\dot{\bf R}\times{\bf B}\Big)\\
M\ddot{\bf R}&= -e\Big[\big({\bf r}\cdot\nabla_{\bf R}\big){\bf E} +\frac{1}{c}\dot{\bf R}\times \big({\bf r}\cdot\nabla_{\bf R} \big){\bf B}\Big]\\
&\ \ \ \ +\frac{e}{c}{\bf B} \times\dot{\bf r}
\label{newtoneqs}
\end{aligned}
\end{equation}
together with the full set of Maxwell's equations with charge and current densities $\rho=e({\bf r}\cdot\nabla)\delta({\bf x}-{\bf R})$ and ${\bf J}=-e\dot{\bf r}\delta({\bf x}-{\bf R})+e\dot{\bf R}({\bf r}\cdot\nabla)\delta({\bf x}-{\bf R})$. Here, $\bf r$ and $\bf R$ are the oscillator's relative and center of mass coordinates with associated effective masses $m$ and $M$ \cite{Cherqui_2014}, respectively, and the fields ${\bf E}({\bf R})$ and ${\bf B}({\bf R})$ are evaluated at the dipole's center of mass position $\bf R$.

Elimination of the current density $\bf J$ from $P$ using Maxwell's equations results in Poynting's theorem for the continuity of energy in the field, i.e., $\dot u+\nabla\cdot{\bf S}=-{\bf E}\cdot{\bf J},$ where $u=(8\pi)^{-1}({\bf E}^2+{\bf B}^2)$ is the energy density and ${\bf S}=(c/4\pi){\bf E}\times{\bf B}$ is the energy flux. Likewise, elimination of $\bf E$ from $P$ using the Newton equations (Eqs. \eqref{newtoneqs}) produces the material energy continuity equation, 
\begin{equation}
\frac{d}{dt}\Big[\frac{1}{2}m\dot{\bf r}^2+\frac{1}{2}m\omega_0^2{\bf r}^2+\frac{1}{2}M\dot{\bf R}^2\Big]=P,
\label{consvenergy}
\end{equation} 
implicitly identifying the (a priori unknown) material Hamiltonian $H=({1}/{2})m\dot{\bf r}^2+({1}/{2})m\omega_0^2{\bf r}^2+({1}/{2})M\dot{\bf R}^2$ from knowledge of only the equations of motion and energy transfer rate.

Together, these continuity equations yield the composite light-matter conservation relation 
\begin{equation}
\frac{d}{dt}\Big[H+\int d{\bf x}\,u\Big]=-\int r^2d\Omega\,\hat{\bf n}\cdot{\bf S},
\label{energy}
\end{equation}
with radiatively open boundary conditions on the surface bounding the integration volume with surface normal vector $\hat{\bf n}.$

\subsubsection{Connection to momentum pseudoscalars}
Given the relation between a harmonic oscillator's linear ${\bf p}_i = \left(im\omega/e\right){\bf d}_i$ and angular ${\bf L}_{ij} = \left(\mathbf{R}_i-\mathbf{R}_j\right)
\times
{\mathbf{p}}_j$ momenta with its dipole moment ${\bf d}_i=-e{\bf r}_i,$ the light-matter chirality ${\cal M}_\Gamma$ of the $\Gamma$th eigenmode can be transformed from the dipole basis to the momentum basis. Specifically, by noticing that ${\bf a}\cdot{\bf b}\times{\bf c}={\bf 0}$ when any pair of vectors is parallel, $\Lambda\text{Re}[d^\Gamma_1(d^\Gamma_2)^*]=-(1/2)(e/m\omega_\Gamma)^2({\mathbf p}^\Gamma_1\cdot{\mathbf L}_{12}^{\Gamma*}+{\mathbf p}^\Gamma_2\cdot{\mathbf L}_{21}^{\Gamma*}+0)$, where $0={\mathbf p}^\Gamma_1\cdot
{\mathbf L}_{21}^{\Gamma*}+{\mathbf p}^\Gamma_2\cdot{\mathbf L}_{12}^{\Gamma*}.$ Thus, ${\cal M}_\Gamma$ may be equivalently expressed in terms of the scalar product of total eigenmode linear ${\bf p}^\Gamma_{\textrm{tot}}={\bf p}_1^\Gamma+{\bf p}_2^\Gamma$ and angular ${\bf L}^\Gamma_{\textrm{tot}} ={\bf L}^\Gamma_{12}+{\bf L}^\Gamma_{21}$ momenta and is rendered axis independent. This result, derived directly from the governing dynamical equations, provides a rigorous foundation for the family of momentum pseudoscalar metrics \cite{PhysRevLett.133.268001,tao2026chiral} and clarifies the roles of inter- (${\bf p}_1^\Gamma\cdot{\bf L}_{12}^{\Gamma*}$ and ${\bf p}_2^\Gamma\cdot{\bf L}_{21}^{\Gamma*}$) versus intra- (${\bf p}_1^\Gamma\cdot{\bf L}_{21}^{\Gamma*}$ and ${\bf p}_2^\Gamma\cdot{\bf L}_{12}^{\Gamma*}$) site contributions \cite{8f2s-rjgy}. 

%Here, it is evident that the addition of the intra-site contributions (i.e., ${\mathbf p}_1\cdot{\mathbf L}_{21}^*+{\mathbf p}_2\cdot{\mathbf L}_{12}^*=0$) renders ${\mathcal{M}}_\Gamma$ axis independent \cite{8f2s-rjgy}. 

For arbitrarily many dipole oscillators of charge $Q_i$ and effective mass $m_i$, ${\cal M}_\Gamma$ generalizes to
%\begin{widetext}
\begin{equation}
\begin{aligned}
 {\mathcal{M}}_\Gamma 
 %&=
%\frac{ik^4_\Gamma}{2r_{12}}\Big(\frac{e}{m\omega_\Gamma}\Big)^2h^{(2)}_1(k_\Gamma r_{ij})
%\Big[{\mathbf p}^\Gamma_1\cdot{\mathbf L}_{12}^{^\Gamma*}+{\mathbf p}^\Gamma_2\cdot{\mathbf L}_{21}^{^\Gamma*}+\underbrace{{\mathbf p}^\Gamma_1\cdot{\mathbf L}_{21}^{\Gamma*}+{\mathbf p}^\Gamma_2\cdot{\mathbf L}_{12}^{\Gamma*}}_{0}\Big] \\
%&=\frac{ik^4_\Gamma}{2r_{12}}\Big(\frac{e}{m\omega_\Gamma}\Big)^2h^{(2)}_1(k_\Gamma r_{ij}){\mathbf p}^\Gamma_{\textrm{tot}}\cdot{\mathbf L}_{\textrm{tot}}^{\Gamma*}\\ 
&=\sum_{i\neq j}\frac{ik_\Gamma^2}{2r_{ij}}\frac{Q_iQ_j}{m_im_jc^2} 
h_1^{(2)}(k_\Gamma r_{ij})\\
&\ \ \ \ \ \ \ \ \times\Big({\bf p}_i^\Gamma\cdot{\bf L}_{ij}^{\Gamma*}+{\bf p}_i^\Gamma\cdot{\bf L}_{ji}^{\Gamma*}\Big),
    \label{MPS2}
   \end{aligned}
    \end{equation}
%\end{widetext}
where the spherical Hankel function $h_1^{(2)}$ originates from the magnetic Green's function of an electric dipole \cite{Jackson2003}. The molecular adaptation of either the real or imaginary part of Eq. \eqref{MPS2} is further generalized to situations where electronic and nuclear motion are adiabatically separated as well as appreciably coupled in Ref. \cite{tao2026chiral}.

%Extension to arbitrarily many pairwise-interacting oscillators is straightforward.

\subsubsection{Equivalent forms of $Z$}
In consideration of a dipole oscillator of charge $-e$ in the presence of the electromagnetic field, the oscillator's relative coordinate  kinetic momentum is ${\bm\pi}\equiv m\dot{\bf r}={\bf p}+(e/c){\bf A}({\bf R}),$ so that $\nabla_{{\bf R}}\times{\bm\pi}=\nabla_{{\bf R}}\times m\dot{\bf r}=(e/c){\bf B}({\bf R}),$ where ${\bf B}({\bf R})=\nabla_{{\bf R}}\times{\bf A}({\bf R})$ is the  magnetic field evaluated at the oscillator's center of mass coordinate ${\bf R}.$ Likewise, the oscillator's center of mass kinetic momentum is ${\bm\Pi}\equiv M\dot{\bf R}={\bf P}+(e/c)({\bf r}\cdot\nabla_{\bf R}){\bf A}({\bf R}),$ so that $\nabla_{{\bf R}}\times{\bm\Pi}=\nabla_{{\bf R}}\times M\dot{\bf R}=(e/c)({\bf r}\cdot\nabla_{\bf R}){\bf B}({\bf R}).$ Given these relations, the matter chirality can be expressed in the following equivalent forms
\begin{equation}
\begin{aligned}
 Z &=m\dot{\bf r}\cdot\nabla_{\bf R}\times\dot{\bf r}+m\dot{\bf R}\cdot({\bf r}\cdot\nabla_{\bf R})\nabla_{\bf R}\times\dot{\bf r}\\
 &=\frac{1}{m}{\bm\pi}\cdot\nabla_{\bf R}\times{\bm\pi}+\frac{1}{M}{\bm\Pi}\cdot\nabla_{\bf R}\times{\bm\Pi}\\
 &=-\frac{1}{c}{\bf j}\cdot{\bf B}({\bf R})-\frac{1}{c}{\bf j}_{\bf R}\cdot({\bf r}\cdot\nabla_{\bf R}){\bf B}({\bf R}),
\end{aligned}
\end{equation}
where ${\bf j}=-e\dot{\bf r}$ and ${\bf j}_{\bf R}=-e\dot{\bf R}.$ Here, the second line makes explicit the connection to the class of generalized helicity metrics for the field $\bf X$ of the form ${\bf X}\cdot\nabla\times{\bf X}$.

\subsubsection{Cross sections for optical and matter chirality}
The optical chirality cross section $\igmas_F^{\textrm{opt}}(\omega)=\langle{\Phi}_F\rangle/|\langle{\bm\varphi}_0\rangle|$ is a measure of the extinction (ext), absorption (abs), and scattering (sca) of optical spin angular momentum from the incident field by a target object under arbitrary incident polarization, where $\langle{\Phi}_F\rangle=-(1/2)\int d{\bf x}{\textrm{Re}[{\bf J}\cdot\nabla\times{\bf E}_F^*]}$ and $\langle{\bm\varphi}_0\rangle=\pm(\omega/c)\langle{\bf S}_0\rangle$ is the incident chiral flux of $\pm$-circularly polarized light proportional to the incident Poynting flux $\langle{\bf S}_0\rangle$. Here, $F=\{\textrm{ext, abs, sca}\}$ with total field ${\bf E}={\bf E}_0+{\bf E}_\textrm{s}$ expressed in terms of its incident (${\bf E}_\textrm{ext}={\bf E}_0$),  total (${\bf E}_\textrm{abs}={\bf E}$), and scattering (${\bf E}_\textrm{sca}=-{\bf E}_\textrm{s}$) components. For linear media, $\langle{\Phi}_{\textrm{ext}}\rangle$ reduces to circular dichroism at specific incident polarization angles and frequencies; however, in general, $\langle{\Phi}_{\textrm{ext}}\rangle$ remains distinct from circular dichroism.

By analogy, the matter chirality cross sections for extinction, absorption, and scattering are defined as $\igmas_F^{\textrm{mat}}(\omega)=\langle Z_F\rangle/|\langle{\bm\varphi}_0\rangle/\omega|,$ where $\langle Z_F\rangle=(1/2\omega)\int d{\bf x}{\textrm{Im}[{\bf J}\cdot\nabla\times{\bf E}_F^*]}.$ Physically, $\langle Z_F\rangle$ enumerates the crossings of the magnetic field ${\bf B}_F=(c/i\omega)\nabla\times{\bf E}_F$ with the magnetic flux loops of each dipole.

\subsubsection{Relation to circular dichroism}
In Ref. \cite{8f2s-rjgy} we showed that the optical chirality extinction ${\igmas}_{\textrm{ext}}^{\textrm{opt}}(\omega)$ reduces to circular dichroism under linearly polarized plane wave excitation at $\phi = \pi/4$ for the BK$_2$ structure. We emphasize that ${\igmas}_{\textrm{ext}}^{\textrm{opt}}(\omega)$ always measures the extinction of canonical spin angular momentum from an arbitrary incident driving field, independent of its wave vector and polarization content, while circular dichroism is limited to measurement of the difference in extinction between left- and right-handed circularly polarized plane waves.

For a general linear scatterer, optical chirality extinction along the angle satisfying
\begin{equation}
\tan\phi_\pm(\omega) = \frac{\Delta T\mp\sqrt{
\Delta T^2+4\text{Re}\{
T_{yx}\}\text{Re}\{
T_{xy}\}}}{2\text{Re}\{
T_{yx}\}}
\end{equation}
is equal to circular dichroism from consideration of the optical theorem \cite{berg2008extinction} and the optical theorem for optical chirality \cite{PhysRevA.92.023813,8f2s-rjgy}. Here ${\bf F}({\bf k}_0,{\bf k}_0) = {\bf T}{\bf E}_0$ is the forward scattering amplitude, expressed in terms of the incident field ${\bf E}_0$, and $\Delta T = \text{Re}\left(T_{yy}-T_{xx}\right)$. Since the entries $T_{ij}$ are dispersive,  ${\igmas}_{\textrm{ext}}^{\textrm{opt}}(\omega)$ for any frequency-independent incident linear polarization and circular dichroism are in general distinct.

\end{document}